\documentclass[pre,twocolumn,showpacs]{revtex4}
\usepackage{amsmath,graphicx}

\begin{document}

\title{Diffusion on a heptagonal lattice}
\author{Seung Ki Baek}
\author{Su Do Yi}
\author{Beom Jun Kim}
\email[Corresponding author, E-mail: ]{beomjun@skku.edu}
\affiliation{Department of Physics, BK21 Physics Research Division, and
Institute of Basic Science, Sungkyunkwan University, Suwon 440-746,
Republic of Korea}

\begin{abstract}
We study the diffusion phenomena on the negatively curved surface made up of
congruent heptagons. Unlike the usual two-dimensional plane, this structure
makes the boundary increase exponentially with the distance from the center,
and hence the displacement of a classical random walker
increases linearly in time. The diffusion of a quantum particle put on the
heptagonal lattice is also studied in the framework of the tight-binding
model Hamiltonian, and we again find the linear diffusion like the classical
random walk. A comparison with diffusion on complex networks is also made.
\end{abstract}


\maketitle


The transport phenomena have been extensively studied on various spatial
structures to reveal the characteristic thermal and electrical properties of
materials. The random walk is one of the classical topics for studying those
phenomena in the nonequilibrium statistical physics~\cite{rudnick},
and the transport on highly disordered systems such as complex networks is also
of recent research interest~\cite{bjkim,jdnoh,mulken}. Since the
average path length is often greatly reduced in complex networks, one can easily
expect that a particle diffuses there faster than on the ordinary regular
lattice structures.

Recently, the surface with a constant negative Gaussian curvature is under
active investigation to study geometrical effects on the critical
phenomena~\cite{shima-belo-sausset,baek}.
A negative Gaussian curvature means that the surface
locally looks like a uniform saddle everywhere, with the resulting surface
getting highly curled as one moves outward~\cite{thurston}. Consequently,
the boundary length of a circle with radius $r$ is given by $2 \pi~\sinh~r$,
rather than by $2 \pi r$~\cite{anderson}, which means in turn that the
characteristic path length
increases only logarithmically with the system size $N$~\cite{baek}.
For a positively curved surface, the system becomes
eventually closed, like a sphere, where the magnitude of curvature has to
approach zero in the thermodynamic limit. Accordingly, 
there is little reason to study this case separately.
In contrast, surfaces of a constant negative Gaussian
curvature can be extended indefinitely, which
makes such a geometry apt for studying novel physical properties.
Furthermore, the development of nanotechnology can make it possible to
construct such a structure in reality, and the physical properties
discovered here are also of practical importance.
There already exist theoretical interests in negatively curved
nanostructures with a very large surface area embedded
in a limited volume~\cite{nano}.

It has been reported that the hyperbolic Brownian motion has a
limited direction~\cite{kendall} with a constant outward drift~\cite{monthus}.
Such nontrivial behaviors are nothing mysterious and directly
related to the exponentially increasing length of a boundary: Since there is
always a larger number of points outside than inside by some constant
proportion, a random walker tends to move outward, which allows the random
walker only a limited direction~\cite{karlsson}.
However, the validity of such a simple geometric understanding
of the fast classical diffusion needs to be taken carefully if one compares
the classical diffusion with the quantum one, which composes the main
motivation of this Brief Report.
%

In this work, we present how to simulate the random walk on such a geometry and
reconfirm the linear classical diffusion for a heptagonal lattice structure.
The quantum diffusion of a particle is then investigated in comparison
both with the classical diffusion behavior and with Ref.~\cite{bjkim} 
in which the quantum diffusion time has been shown to
scale as $\tau_q \sim N^{\alpha-1}$ while $\tau_c \sim N^\alpha$ for the
classical diffusion. Also in Ref.~\cite{mulken}, quantum mechanical transport
on graphs has been shown to be faster than the classical one, except on some
finite treelike graphs. In comparison, $\tau_q \sim \tau_c$ is revealed in the
present work on the heptagonal lattice geometry. It is remarkable that one can
drastically accelerate the transport by introducing heptagonal plaquettes to
the lattice~\cite{nano}.

\begin{figure}
\includegraphics[width=0.46\textwidth]{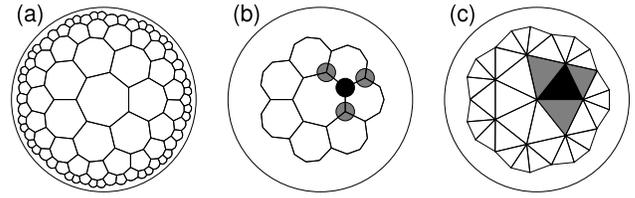}
\caption{(a) Heptagonal lattice represented on the Poincar{\'e} disk
with the total number of levels $L=4$. The metric is given in such a way that
all heptagons are congruent and the circle indicates the points at infinity.
(b) Random walk on the heptagonal lattice where a walker at the black
dot has three choices (colored in gray) at each time. (c) In the dual
lattice, walking around a heptagon is equivalent to the 
reflection of a triangle.}
\label{fig:walker}
\end{figure}

Let us describe the structure of the heptagonal lattice.
Similarly to the two-dimensional (2D) plane which can be covered by
congruent triangles, squares, and hexagons,
a negatively curved surface can be tiled by polygons through
the hyperbolic tessellation~\cite{coxeter}: Suppose
a lattice where $q$ regular $p$-gons meet at each vertex. For example, the
triangular lattice is denoted as $p=3$ and $q=6$. It is known that the
tiling with given $p$ and $q$ covers the negatively curved surface if
$(p-2)(q-2)>4$. If we choose its basic element as the simplest polygon, the
regular triangle, at least seven triangles should meet at each vertex
leading to $\{p,q\}=\{3,7\}$. Taking its dual lattice, we
obtain the heptagonal lattice structure with $\{p,q\}=\{7,3\}$.
The resulting structure can be most suitably represented in the
Poincar{\'e} disk~\cite{green} as in Fig.~\ref{fig:walker}(a).
Since the heptagons are located in a concentric fashion,
we denote the central heptagon as the first level. Accordingly, its
seven nearest-neighbor heptagons constitute the second level, and the next
nearest ones the third level, and so on. Figure~\ref{fig:walker}(a) thus describes
the heptagonal lattice up to the level $l=4$. Let $H(l)$ be the set of
heptagons in the $l$th level. The set of vertices on the outward boundary
of $H(l)$ is called the $l$th layer.
As one can see, this structure uniformly and completely fills the surface
and provides an adequate tool for studying physics
of the negative Gaussian curvature.
For numerical calculations, we may use the heptagonal lattice constructed
up to some finite level, $L$, and capture the finite size effects
by varying $L$.

\begin{figure}
\includegraphics[width=0.46\textwidth]{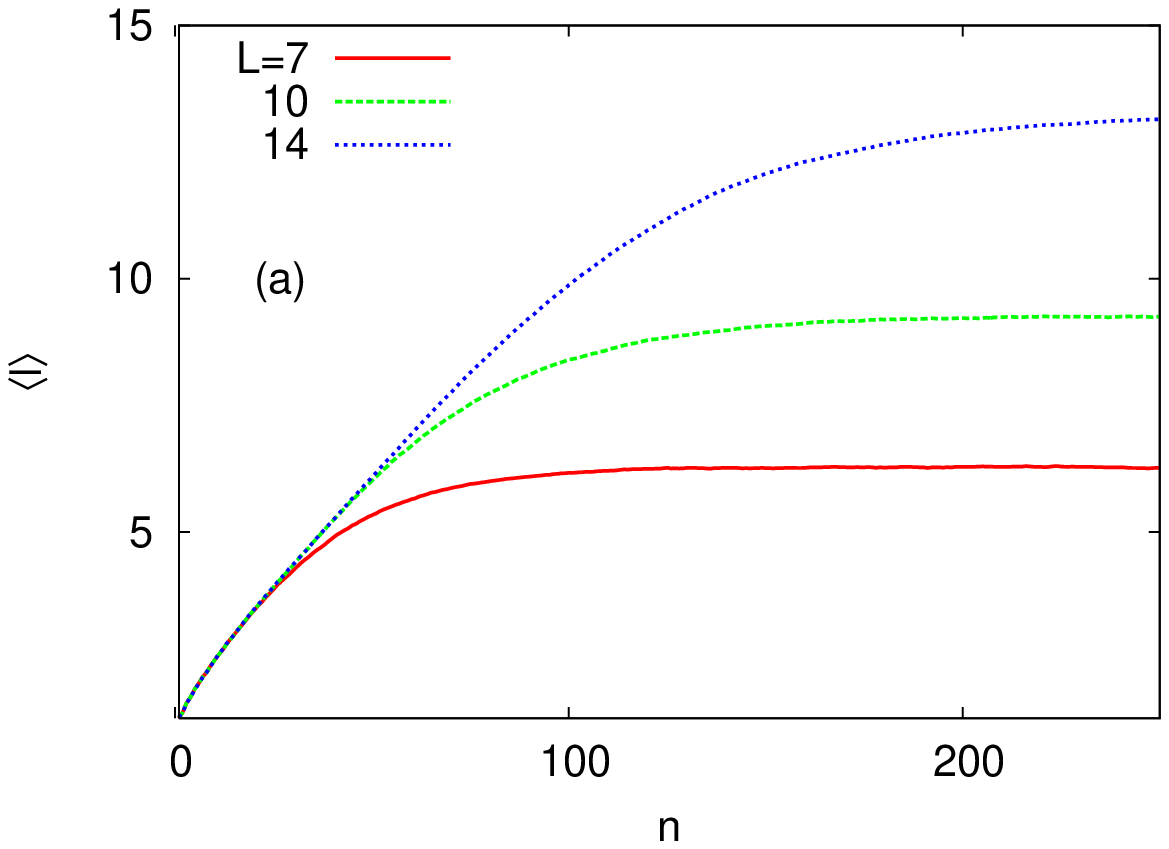}
\includegraphics[width=0.46\textwidth]{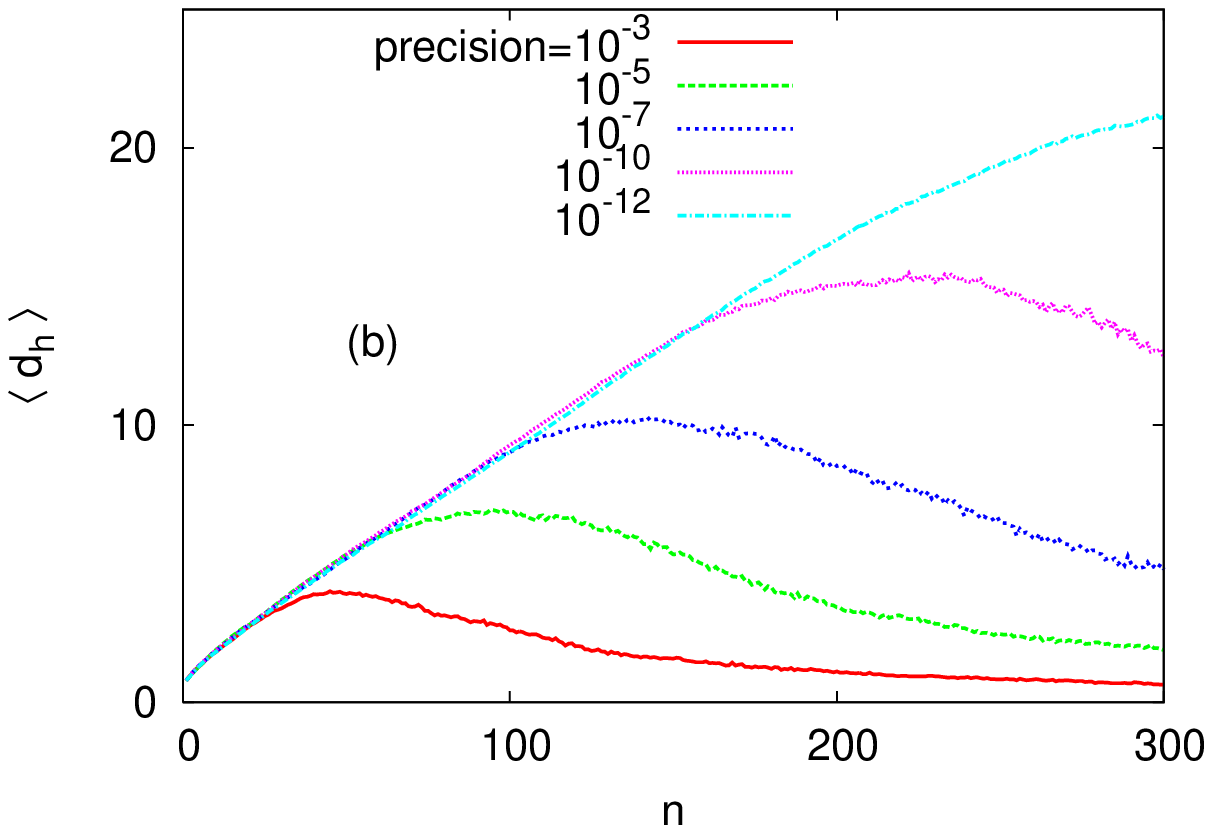}
\caption{(Color online) (a) Average level $\langle l \rangle$
of the random walker versus the time $n$. The heptagonal
lattice of the size $L$ is first built, and we perform
random walks as in Fig.~\ref{fig:walker}(b). We see $\langle l \rangle
\propto n$ for small $n$, while $\langle l \rangle$
eventually saturates as the walker approaches the external boundary.
(b) Average distance $\left< d_h \right>$ from the origin 
versus the time $n$, obtained from the method in Fig.~\ref{fig:walker}(c). 
The deviation from
the linear behavior is not due to the finiteness of the lattice
as in (a), but due to accumulated numerical errors.
All the averages were taken from $10^3$ independent realizations.}
\label{fig:dist}
\end{figure}

We first study numerically the random walk problem in our heptagonal
lattice structure, and the results are compared with
numerical solutions of the diffusion equation for the hyperbolic
geometry.
A possible way of simulating a random walker on the heptagonal lattice
is to identify the connection structure first and then let a particle move
along the edges. In Fig.~\ref{fig:dist}(a), the position of the
random walker measured by the level $l$ is shown as a function of the 
time step $n$ for the
heptagonal lattices of sizes $L=7$, 10, and 14. It is clearly shown
that the random walker drifts away from the starting position $l=1$ linearly
in time. As the lattice becomes larger (as $L$ is increased), the linear
diffusion regime becomes extended, indicating 
$\langle l \rangle \propto n$ in the thermodynamic limit.
One drawback of this approach is that the number of points 
increases exponentially as we add concentric
layers. The memory constraint restricts the distance from the
origin, and therefore the results severely suffer from finiteness of the
model system under consideration.

A better alternative is obtained from the hyperbolic tessellation. Since the
heptagonal lattice is dual to $\{3,7\}$, the transition between the
neighboring points coincides with reflecting a hyperbolic triangle
[see Fig.~\ref{fig:walker}(c)].
Since the Poincar{\'e} disk can be identified with a unit disk on the
complex plane, we start from a regular triangle $(z_1, z_2, z_3)$ where
$z_i$'s are complex numbers. Let one of the vertices, say $z_1$, be located
at the origin. Setting the interior angle at that point to be $2\pi/7$, the
triangle is regular only when $|z_1-z_2| = |z_1-z_3| \approx 0.496 97$,
as this surface has its own intrinsic length scale, i.e., the
curvature~\cite{green}.
If a triangle $(z_i, z_j, z_k)$ is reflected around the edge along $z_i$ and
$z_j$, we get a new triangle $(z_i, z_j, z_k')$ with
$z_k' = w + \xi^2/(\overline{z_k}-\overline{w})$, 
where 
$w=(|z_i|^2z_j - z_i|z_j|^2-z_i + z_j)/(\overline{z_i}z_j - z_i\overline{z_j})$, $\xi=|w-z_i|=|w-z_j|$, and $\overline{z}$ is the complex conjugate of $z$.
The center of a triangle, $z$, has the hyperbolic distance from the origin by
$d_h(z) = \ln [(1+|z|)/(1-|z|)]$. Note that $d_h$ diverges to infinity as
$|z| \rightarrow 1$, which is consistent with the definition of the Poincar{\'e} disk.
The result is depicted in Fig.~\ref{fig:dist}(b), which clearly shows 
that the distance is linearly proportional to time. 
The numerical inaccuracy becomes larger as we repeat the reflection
of triangles. In order to confirm the origin of the deviation from the 
linear behavior $\langle d_h \rangle \sim n$  at large $n$, we
intentionally assign numerical precision and observe how $\langle d_h \rangle$
depends on it. In Fig.~\ref{fig:dist}(b), it is seen that as we use
better numerical precision, the linear regime becomes more extended, indicating
that the deviation from the linear behavior originates from the simple
artifact of the numerical accuracy. Beyond the deviation point of
each curve in Fig.~\ref{fig:dist}, the numerical values are not
trustworthy and thus the decrease of $\langle d_h \rangle$ 
should not be taken as real.

\begin{figure}
\includegraphics[width=0.46\textwidth]{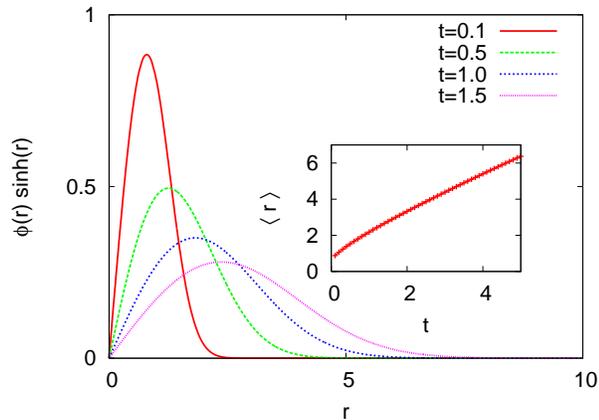}
\caption{(Color online) The solution $\phi(r,t)$ of the 
hyperbolic diffusion equation.
Inset: Average distance from the origin 
$\left< r \right> = \int r \phi(r) \sinh~r ~dr$ increases linearly
in time.}
\label{fig:dif}
\end{figure}

For the random walk on a 2D flat surface,
the walker has zero probability of escape by P{\'o}lya's theorem
which states that the walk becomes transient for dimensions larger than 
two~\cite{rudnick}.
If we denote the probability density of finding a particle at a position ${\bf r}$
after time $t$ as $\phi({\bf r},t)$, the diffusion process is
described by
${\partial \phi}/{\partial t} = \nabla^2 \phi$.
Solving this equation by Fourier transformation, one can see that the
expected displacement from the origin is proportional to the square root of
time in the 2D flat surface. 
The Laplace operator $\nabla^2$ is changed to the Laplace-Beltrami 
operator for the hyperbolic metric~\cite{young},
\begin{equation}
\triangle = 
\frac{1}{\sinh~r} \frac{\partial}{\partial r} \left( \sinh~r
\frac{\partial  }{\partial r} \right) + \frac{1}{\sinh^2~r} \frac{\partial^2
}{\partial \theta^2},
\end{equation}
and we get the solution of the hyperbolic diffusion equation, 
$\partial \phi/\partial t = \triangle \phi$~\cite{monthus,banica}:
\begin{equation}
\phi(r,\theta;t) \propto \frac{e^{-t/4}}{t^{3/2}} \int\int dr'
d\theta' \phi(r', \theta';0) I(t,\rho) \sinh~r' 
\label{eq:sol}
\end{equation}
with the distance $\rho$ between two positions
$(r,\theta)$ and $(r', \theta')$, and
$I(t,\rho) \equiv \int_{\rho}^{\infty} (s e^{-s^2/4t}/\sqrt{\cosh~s - \cosh~\rho})ds$.
The numerical integration is then performed to get the result
presented in Fig.~\ref{fig:dif}, which again shows the linear
diffusion $\langle r \rangle \propto t$. 
We thus conclude that our different approaches, i.e., discrete random walks
by triangle reflections on the Poincar{\'e} disk and the solution of the 
hyperbolic diffusion equation, unanimously confirm that 
the distance from the origin increases linearly in time.
In comparison to $\langle d \rangle \sim \sqrt{t}$
for the 2D flat surface, the diffusion occurs much faster
in the negatively curved surface.

\begin{figure}
\includegraphics[width=0.46\textwidth]{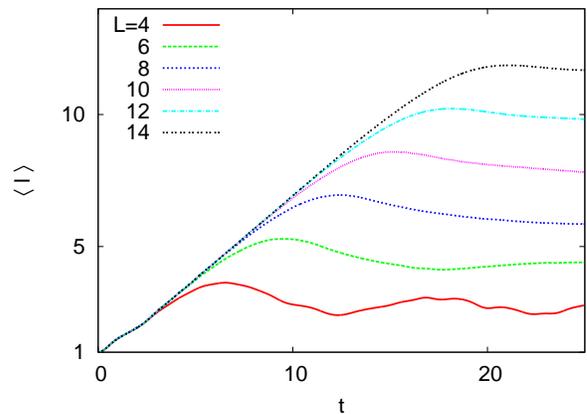}
\caption{(Color online) Diffusion of the tight-binding quantum
particle  on the heptagonal lattice. As the time $t$ evolves,
the quantum particle diffuses toward the upper layers. Note that
the linear regime extends as the size becomes larger.}
\label{fig:level}
\end{figure}

We next study the diffusion of a tight-binding quantum particle
in the heptagonal lattice.
The quantum diffusion phenomena,  and the existence and nature
of the localization transition have been studied on regular lattice
structures for a long time. Motivated by the intensive research interest
of complex networks, recent years have observed the beginning
of research on quantum mechanical systems put on the structure
of complex networks~\cite{bjkim,mulken,cpzhu-giraud}.
We assume that the wave function is localized on the first layer containing
seven points [see Fig.~\ref{fig:walker}(a)].
The time evolution of the quantum particle is governed by
the Schr{\"o}dinger equation 
$ i (\partial  / \partial t) |\psi\rangle = {\cal H} |\psi\rangle $
($\hbar \equiv 1$). 
By using the perfectly localized states as basis kets,
we apply the tight-binding approximation that ${\cal H}_{ij} = 1 (0)$ if 
two points $i$ and $j$ are connected (not connected).
The Schr{\"o}dinger equation is then numerically integrated by the fourth
order Runge-Kutta method. The spreading of the wave packet
is measured by the average layer, $\left< l \right>$, depicted in
Fig.~\ref{fig:level}. Again, one can see clearly that the particle 
diffuses linearly in time as in the classical diffusion. 
The saturation comes from the finiteness of our lattice.
It is to be noted that
in the negatively curved heptagonal lattice, both the quantum and
the classical diffusion exhibit the linear diffusion.

\begin{figure}
\includegraphics[width=0.46\textwidth]{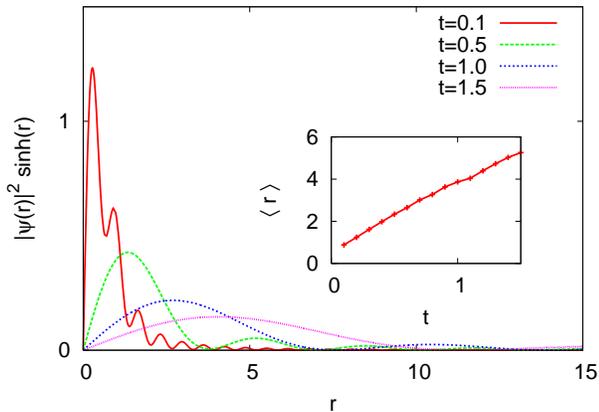}
\caption{(Color online) Solution of the quantum free particle Schr{\"o}dinger
equation in the negatively curved surface.
Inset: Average distance from the origin 
$\left< r \right> = \int r |\Psi(r)|^2 \sinh~r ~dr$ is shown
to increase linearly in time.}
\label{fig:sch}
\end{figure}

We then take an alternative approach to investigate the quantum diffusion
problem, through the use of the mapping of the above tight-binding
Hamiltonian to the one for a free particle. We note that the shift
of energy by a constant amount does not change any measurable
quantity and makes the transformation ${\cal H} \rightarrow {\cal H} - k I$
with a degree $k$ (the number of neighbors, e.g., $k = 3$ for
the heptagonal lattice) and the identity operator $I$.
This simple transformation makes the Hamiltonian proportional to
the lattice Laplacian,  except for the outermost points where $k \neq 3$.
In short, the tight-binding Hamiltonian can be phenomenologically treated
as that of the free particle simulated on the discrete lattice,
described by $i \partial \Psi / \partial t =  \triangle \Psi$
with the wave function $\Psi = \langle {\bf r}| \psi \rangle$.
Accordingly,
if we substitute the time $t$ in the Schr{\"o}dinger equation 
by the imaginary time $i t$, it takes exactly the same form as
the diffusion equation. It is then straightforward to apply
the previous result on classical diffusion in hyperbolic lattice
(see Fig.~\ref{fig:dif}) to obtain the propagating solution of
the quantum particle on the hyperbolic plane~\cite{banica} 
(see Fig.~\ref{fig:sch}). It is revealed that 
the average distance again exhibits the linear diffusion 
property as shown in the inset of Fig.~\ref{fig:sch} in accordance with 
Fig.~\ref{fig:level}.

In the Euclidean or disordered structures, the quantum diffusion has
been known to be faster than the classical one~\cite{bjkim,mulken}. 
In contrast, the classical
motion now becomes comparable to the quantum one in speed on the negatively
curved surface, with the aid of
the geometrical drift. Some differences, however, seem to remain: The
geometrical drift pushes the particle only in the outward direction from the
starting point. Based on the report for tree structures, a quantum
particle is believed to readily propagate back to the origin~\cite{childs}.
In fact, a classically diffusing particle acts somewhat like
a flow, even if the initial direction may be chosen randomly, in the sense
that it appears to have some nonzero velocity. If one
introduces a real flow here, however, the flux would soon become negligible
because of the exponentially increasing boundary. 
Our observation is qualitatively similar to the report for the Bethe
lattice~\cite{monthus}, since the angular movement is effectively
suppressed in the long run. Yet it is still notable that the heptagonal
lattice provides a better representation for the negatively curved surface,
and that
the numerical simulation could be taken with more ease by the hyperbolic
tessellation technique explained above.
One may expect that the transport can be enhanced by
introducing a negative Gaussian curvature in biological
or engineering applications.

In summary, we have investigated the classical and quantum diffusions in the
heptagonal lattice, which is a discrete representation of the surface with a constant
negative curvature. Even a classical particle has been shown to diffuse so fast
that the average displacement is linearly proportional to time. The
quantum diffusion has also been shown to exhibit the same linear diffusion
behavior. Those results on discrete lattices were also
confirmed in a continuum by solving a hyperbolic diffusion equation
in real and imaginary times for classical and quantum diffusion, respectively.

\acknowledgments

This work was supported by the Korea Research Foundation Grant funded by the
Korean Government (MOEHRD) with the Grant No. KRF-2005-005-J11903 (S.K.B.)
and KRF-2006-312-C00548 (B.J.K.).

\bibliographystyle{revtex}

\begin{thebibliography}{10}
\providecommand*{\bibinfo}[2]{#2}
\providecommand*{\eprint}[1]{#1}
\providecommand*{\url}[1]{#1}
\bibitem{rudnick}
\bibinfo{author}{J.~Rudnick} and \bibinfo{author}{G.~Gaspari},
  \bibinfo{title}{\emph{Elements of the Random Walk: An Introduction for
  Advanced Students and Researchers}} (\bibinfo{publisher}{Cambridge University
  Press}, New York, \bibinfo{year}{2004}).
\bibitem{bjkim}
\bibinfo{author}{B.~J. Kim}, \bibinfo{author}{H.~Hong}, and
  \bibinfo{author}{M.~Y. Choi}, \bibinfo{journal}{Phys. Rev. B}
  \bibinfo{volume}{\textbf{68}}, \bibinfo{pages}{014304}
  (\bibinfo{date}{2003}).
\bibitem{mulken} O. M{\"u}lken and A. Blumen, Phys. Rev. E {\bf 73}, 066117 (2006).
\bibitem{jdnoh}
\bibinfo{author}{J.~D. Noh} and \bibinfo{author}{H.~Rieger},
  \bibinfo{journal}{Phys. Rev. Lett.} \bibinfo{volume}{\textbf{92}},
  \bibinfo{pages}{118701} (\bibinfo{date}{2004});
\bibinfo{author}{J.~D. Noh} and \bibinfo{author}{S.-W. Kim},
  \bibinfo{journal}{J. Korean Phys. Soc.} \bibinfo{volume}{\textbf{48}},
  \bibinfo{pages}{S202} (\bibinfo{date}{2006}).
\bibitem{shima-belo-sausset}
\bibinfo{author}{H.~Shima} and \bibinfo{author}{Y.~Sakaniwa},
  \bibinfo{journal}{J. Stat. Mech. (2006) P08017};
  \bibinfo{journal}{J. Phys. A: Math. Theor.} \bibinfo{volume}{\textbf{39}},
  \bibinfo{pages}{4921} (\bibinfo{date}{2006});
\bibinfo{author}{L.~R.~A. Belo}, \bibinfo{author}{N.~M. Oliveira-Neto},
  \bibinfo{author}{W.~A. Moura-Melo}, \bibinfo{author}{A.~R. Pereira}, and
  \bibinfo{author}{E.~Ercolessi}, \bibinfo{journal}{Phys. Lett. A}
  \bibinfo{volume}{\textbf{365}}, \bibinfo{pages}{463} (\bibinfo{date}{2007});
\bibinfo{author}{F.~Sausset} and \bibinfo{author}{G.~Tarjus},
  \bibinfo{journal}{J. Phys. A: Math. Theor.} \bibinfo{volume}{\textbf{40}},
    \bibinfo{pages}{12873} (\bibinfo{date}{2007}).
\bibitem{baek}
\bibinfo{author}{S.~K. Baek} and \bibinfo{author}{B.~J. Kim},
  \bibinfo{journal}{Europhys. Lett.} \bibinfo{volume}{\textbf{79}},
  \bibinfo{pages}{26002} (\bibinfo{date}{2007}).
\bibitem{thurston}
\bibinfo{author}{W.~P. Thurston} and \bibinfo{author}{J.~R. Weeks},
  \bibinfo{journal}{Sci. Am.} \bibinfo{volume}{\textbf{251}},
  \bibinfo{pages}{108--120}
  (\bibinfo{date}{1984}).
\bibitem{anderson}
\bibinfo{author}{J.~W. Anderson}, \bibinfo{title}{\emph{Hyperbolic Geometry}}
  (\bibinfo{publisher}{Springer-Verlag}, London, \bibinfo{year}{1999}).
\bibitem{nano} D. Vanderbilt and J. Tersoff, Phys. Rev. Lett. {\bf 68},
511 (1992); N. Park, M. Yoon, S. Berber, J. Ihm,  E. Osawa, and D.
Tom{\'a}nek, {\it ibid.} {\bf 91}, 237204 (2003).
\bibitem{kendall}
\bibinfo{author}{W.~S. Kendall}, \bibinfo{journal}{S{\'e}m.
  Probab. (Strasbourg)} \bibinfo{volume}{\textbf{18}},
  \bibinfo{pages}{70} (\bibinfo{date}{1984}).
\bibitem{monthus}
\bibinfo{author}{C.~Monthus} and \bibinfo{author}{C.~Texier},
  \bibinfo{journal}{J. Phys. A: Math. Theor.} \bibinfo{volume}{\textbf{29}},
  \bibinfo{pages}{2399} (\bibinfo{date}{1996}).
\bibitem{karlsson}
\bibinfo{author}{A.~Karlsson}, in \emph{Proceedings of a Workshop at the
  Schr{\"o}dinger Institute, Vienna, 2001} (\bibinfo{publisher}{de Gruyter},
  Berlin, \bibinfo{year}{2004}), \bibinfo{pages}{p. 459}.
\bibitem{coxeter}
\bibinfo{author}{H.~S.~M. Coxeter}, \bibinfo{journal}{Bull. Can. Math.}
  \bibinfo{volume}{\textbf{40}}, \bibinfo{pages}{158} (\bibinfo{date}{1997}).
\bibitem{green}
\bibinfo{author}{M.~J. Greenberg}, \bibinfo{title}{\emph{Euclidean and
  Non-Euclidean Geometries: Development and History}} (\bibinfo{publisher}{W.
  H. Freeman}, New York, \bibinfo{year}{1993}), 3rd ed.
\bibitem{young}
\bibinfo{author}{E.~C. Young}, \bibinfo{title}{\emph{Vector and Tensor
  Analysis}}, 2nd ed. (\bibinfo{publisher}{Marcel Dekker}, New York,
  \bibinfo{year}{1993}).
\bibitem{banica}
\bibinfo{author}{V.~Banica}, 
  \bibinfo{journal}{Commun. Partial Differ. Equ.}
  \bibinfo{volume}{\textbf{32}}, \bibinfo{pages}{1643} (\bibinfo{date}{2007}).
\bibitem{cpzhu-giraud} C.-P. Zhu and S.-J. Xiong, Phys. Rev. B {\bf 62}, 14780
(2000); O. Giraud, B. Georgeot, and D. L. Shepelyansky, Phys. Rev. E {\bf 72},
036203 (2005).
\bibitem{childs}
\bibinfo{author}{A.~M. Childs}, \bibinfo{author}{E.~Farhi}, and
  \bibinfo{author}{S.~Gutmann}, \bibinfo{journal}{Quantum Inf.
  Process.} \bibinfo{volume}{\textbf{1}}, \bibinfo{pages}{35}
  (\bibinfo{date}{2002}).
\end{thebibliography}

\end{document}